\documentclass[aps,prl,nofootinbib,superscriptaddress,notitlepage,preprintnumbers,twocolumn,reprint]{revtex4-1}

\usepackage{graphicx}
\usepackage{amsmath,graphicx,verbatim,epsfig}
\usepackage{epstopdf}
\usepackage[utf8x]{inputenc}
\usepackage[colorlinks=true,linkcolor=blue,citecolor=blue]{hyperref}
\usepackage[usenames,dvipsnames]{color}
\usepackage{bm}
\usepackage{cancel}

\newcommand{\Li}{\text{Li}}

\newcommand{\eps}{\varepsilon}

\newcommand{\nn}{\nonumber}

\newcommand{\bn}{{\bar n}}

\newcommand{\nslash}{{\not \!n}}

\newcommand{\be}{\begin{equation}}
\newcommand{\ee}{\end{equation}}
\newcommand{\bea}{\begin{eqnarray}}
\newcommand{\eea}{\end{eqnarray}}
\newcommand{\balign}{\begin{align}}
\newcommand{\ealign}{\end{align}}

\newcommand{\sandwich}[3]{\left< #1 \right | #2 \left | #3 \right >}

\newcommand{\bg}{\begin{gather}}
\newcommand{\foma}{\end{gather}}

\newcommand{\noopsort}[1]{}

\newcommand{\vecb}[1]{\mbox{\boldmath $#1$}}
\newcommand{\vecbe}[1]{\mbox{\boldmath ${\scriptstyle #1}$}}

\def\z{\zeta}

\def\<{\langle}
\def\>{\rangle}

\def\g{\gamma}  \def\G{\Gamma}
\def\d{\delta}
\def\D{\Delta}
\def\l{\lambda}

\def\m{\mu}
\def\n{\nu}

\def\z{\zeta}

\def\({\left(}
\def\[{\left[}
\def\){\right)}
\def\]{\right]}

\def\ln{\hbox{ln}}

\def\nslash{n\!\!\!\slash}

\def \le { \left    }
\def \ri { \right }

\def\bp{\bar p}

\newcommand{\eq}[1]{Eq.~\eqref{#1}}

\begin{document}
\preprint{NIKHEF 2015-028}
\title{The Transverse Momentum Dependent Fragmentation Function at NNLO}

\author{Miguel G. Echevarria}
\email[]{m.g.echevarria@nikhef.nl}
 \affiliation{Nikhef Theory Group, Science Park 105, 1098XG Amsterdam, the Netherlands}
\affiliation{Department of Physics and Astronomy, VU University Amsterdam, De Boelelaan 1081, NL-1081 HV Amsterdam, the Netherlands}

\author{Ignazio Scimemi}
\email[]{ignazios@fis.ucm.es}
\affiliation{Departamento de F\'isica Te\'orica II, Facultad de Ciencias F\'sicas, Universidad Complutense de Madrid, Spain}

\author{Alexey Vladimirov}
\email[]{vladimirov.aleksey@gmail.com} \affiliation{Department of Astronomy and Theoretical Physics, Lund University, Solvegatan 14A, SE 223 62,
Lund, Sweden} \affiliation{Institut f\"ur Theoretische Physik, Universit\"at Regensburg, D-93040 Regensburg, Germany}

\date{\today}

\begin{abstract}

We calculate the unpolarized transverse momentum dependent fragmentation function (TMDFF) at next-to-next-to-leading order (NNLO), evaluating
separately TMD soft factor and TMD collinear correlator. 
For the first time the cancellation of spurious rapidity divergences in a properly defined individual TMD beyond the first non-trivial order is shown. 
This represents a strong check of the given TMD definition.
We extract the matching coefficient necessary to perform the transverse momentum resummation at next-to-next-to-next-to-leading-logarithmic
accuracy. 
The universal character of the soft function, which enters the definition of all (un)polarized TMD distribution/fragmentation functions, facilitates the future calculation of all the other TMDs and their coefficients at NNLO, pushing forward the accuracy of theoretical predictions for the current and next generation of high energy colliders.


\end{abstract}


\maketitle

{\bf Introduction.} 
Multi-differential cross sections play a central role in our understanding of QCD dynamics. 
In this context the definition of transverse momentum dependent functions (TMDs) has been recently revisited, updating the pioneering work of Collins and Soper \cite{Collins:1981uw,Collins:1981uk}, in order to solve the subtle issue of the cancellation of spurious rapidity divergences inside an individual TMD. 
As a result, one has achieved the factorization theorems for Drell-Yan, Vector Boson/Higgs Production, Semi-Inclusive Deep Inelastic Scattering (SIDIS) and $e^+e^-\rightarrow 2\; hadrons$ processes in terms of individually well-defined TMDs \cite{Collins:2011zzd,Echevarria:2012js,Echevarria:2014rua}. 
All these processes are fundamental for current high energy colliders, like the LHC, KEK, SLAC, JLab or RHIC, and future planned facilities, like the EIC, AFTER, the LHeC or the ILC.

While the formulation of the factorization theorems is solid, a direct evaluation of an individual TMD at two loops is still lacking.  
Such a calculation provides a fundamental check of the factorization theorem and important information for data analysis.
The one-loop TMDs with various quantum numbers have been computed by several groups \cite{Echevarria:2014rua,Aybat:2011zv,GarciaEchevarria:2011rb,Vladimirov:2014aja,Bacchetta:2013pqa,Echevarria:2015uaa,Zhu:2013yxa}.
At two loops some properties of TMDs have been deduced from cross section calculations carried out in QCD (see e.g. \cite{Catani:2011kr,Catani:2012qa,Catani:2013tia,Gehrmann:2012ze,Gehrmann:2014yya}).
In this work we present the result of the next-to-next-to-leading-order (NNLO) evaluation of the unpolarized transverse momentum dependent fragmentation function (TMDFF), which is an essential part of TMD factorization theorems, both in the non-singlet and singlet channels.

The evaluation of individual TMDs at higher orders is an utterly non-trivial check for their definition, since starting from the two-loop order, the singularities of various types mix up. 
The two-loop calculation (for the first time presented in this letter) shows that the combination of factors disentangles and cancels the spurious rapidity divergences within the proper definition of one TMD.
The intermediate pieces of the calculation are also relevant per se. 
In fact we define a regulator for rapidity divergences which can be used in combination with standard dimensional regularization for evaluation of any TMDs.   The Soft Function (which is  essential for the TMD definition and whose two-loop result will be presented in a forthcoming publication~\cite{Echevarria:2015byo}) is a key element for the  NNLO calculation of all (polarized) TMDs. 

In this letter we present the matching coefficient of the unpolarized quark TMDFF onto the integrated fragmentation function (FF) at NNLO (both non-singlet and singlet channels), using explicitly the formalism of Ref.~\cite{Echevarria:2014rua}. 
This result can be immediately used in forthcoming phenomenological applications, see e.g. Ref.~\cite{Bacchetta:2015ora,Angeles-Martinez:2015sea}.  
Our consideration fills the gap in TMD phenomenology, because NNLO coefficients for Transverse Momentum Dependent Parton Distribution Function (TMDPDF), another important ingredient of TMD factorization theorems, can be extracted from the NNLO calculations made in the related formalism \cite{Catani:2011kr,Catani:2012qa,Catani:2013tia,Gehrmann:2012ze,Gehrmann:2014yya}.

{\bf Definitions.}
Individually TMDs are defined as a product of two separate matrix elements, the (square root of the) Soft Function (SF) and the corresponding collinear matrix element. 
The SF is a spin and process independent vacuum expectation value of Wilson lines
\cite{Collins:2011zzd,Echevarria:2012js}, and it is defined as
\begin{widetext}
\begin{eqnarray} \label{eq:softf}
S(\vecb k_{s\perp}) &=& 
\int\frac{d^2\vecb b_\perp}{(2\pi)^2} e^{i\vecbe b _\perp \cdot \vecbe k _{s\perp}} 
\frac{1}{N_c} 
\sandwich{0}{{\rm Tr}\, \le[S_n^{T\dagger} \tilde S_\bn^T \ri](0^+,0^-,\vecb b_\perp)
\le[\tilde S^{T\dagger}_\bn S_n^T\ri](0)}{0},
\end{eqnarray}
where $S_n$ and ${\tilde S}_\bn$ stand for soft Wilson lines along light-cone directions\footnote{The
superscript $T$ on Wilson lines in \eq{eq:softf} implies subsidiary transverse links from the light-cone infinities to transverse infinity,
see details in Refs.~\cite{Belitsky:2002sm,Idilbi:2010im,GarciaEchevarria:2011md}. These links guaranty gauge invariance and are necessary for
calculations in singular gauges. 
The presented calculation has been performed in Feynman gauge, where the contribution of transverse links vanishes.},
and $n$ and $\bn$ are light-cone vectors ($n^2=\bn^2=0,\; n\cdot\bn=2$). 
The unpolarized collinear matrix element is defined as
\begin{eqnarray}\label{eq:pure_collinear_def}
\D_{i\to h}^{(0)}(z,\vecb{\hat P}_{h\perp})&=&
\frac{1}{2z} \int\frac{dy^+d^2\vecb b_\perp}{(2\pi)^3} e^{i(y^+k_\bn^--\vecbe b _\perp \cdot
\vecbe k _{\bn\perp})}
{\rm Tr}
\sum_{\,\,X}\!\!\!\!\!\!\!\!\int \sandwich{0}
{\le[\tilde W_\bn^{T\dagger} \psi\ri](y^+\!\!,0^-\!\!,\vecb b_\perp)}{X;P_h}
\nslash \sandwich{X;P_h}{\le[\bar \psi \tilde W^T_{\bn }\ri](0)}{0}\Bigg|_{\text{zb}},
\end{eqnarray}
\end{widetext}
where index $i$ refers to the parton flavor, $P_h$ is the hadron momentum, $z$ is the Bjorken variable, $k_\bn^-={\hat P}_h^{-}/z$ and $\vecb
k_{n\perp}=-\vecb {\hat P}_{h\perp}/z$. 
The subscript ``zb'' stands for zero-bin subtracted\footnote{\label{zb_footnote}
The zero-bin subtraction is the term used in the Soft Collinear Effective Theory (SCET) literature to account for the double counting with the soft sector.
These definitions are equivalently stated in QCD and SCET, see
e.g. Refs.~\cite{Lee:2006nr,Idilbi:2007ff,Idilbi:2007yi,Echevarria:2012js,Collins:2012uy}. 
Here, we apply this term since the definition that we follow \cite{Echevarria:2014rua}
was originated within SCET. 
However, the present calculation  is performed in standard QCD. 
For the used regularization, the application of zero-bin subtraction is equivalent to calculate the colliner matrix element naively
(\eq{eq:pure_collinear_def}) and then subtract the soft function matrix element in Eq.~(\ref{eq:softf}), thereby obtaining the so-called ``pure
collinear'' matrix element: $\Delta_{\text{pure}}\sim \Delta_{\text{naive}}S^{-1}$. 
The precise details on the definition will be presented in Ref.~\cite{Echevarria:2016scs}.}.
See Ref.~\cite{Echevarria:2014rua} for more details regarding the particular definition of Wilson lines.

Individually both matrix elements have rapidity divergences at every order in the perturbative expansion. 
These divergences are neither ultraviolet (UV) nor long-distance ones and, in principle, are not sensitive to confining dynamics \cite{Collins:2011zzd,GarciaEchevarria:2011rb,Echevarria:2012js,Chiu:2012ir}. 
As argued in Refs.~\cite{Collins:2011zzd,GarciaEchevarria:2011rb,Echevarria:2012js}, such divergences can be removed in the correct combination of soft and collinear matrix element.

The essential property of the SF which allows one to remove the rapidity divergences, is that the logarithm of the SF is maximally linear in the logarithmical rapidity divergences. 
Therefore, it can be split into two pieces \cite{Echevarria:2012js},
\begin{align}\label{eq:splitting}&
\tilde{S}({\mathbf L}_\m,{\mathbf L}_{\sqrt{\d^+\d^-}}) = 
\tilde{S}^\frac{1}{2}({\mathbf L}_\m,{\mathbf L}_{\d^+/\n})\,
\tilde{S}^\frac{1}{2}({\mathbf L}_\m,{\mathbf L}_{\n\d^-})
\,,
\end{align}
where $\n$ is an arbitrary, dimensionless and positive real number that transforms as $p^+$ under boosts and 
\footnote{We denote by $p^+$ and $\bp^-$ the large components of the incoming and outgoing parton momenta, respectively, in a SIDIS hard process.}
we introduce the convenient notation
$${\mathbf L}_{X}\equiv\ln(X^2 \vecb b^2 e^{2\gamma_E}/4).$$
Variables $\d^\pm $ are rapidity regulators that one uses in the $n$- and $\bn$-collinear sectors  (our implementation of it is specified later
in Eq.~(\ref{eq:reg})).
Tildes mark quantities calculated in the coordinate space. 
In our calculation the relation in \eq{eq:splitting} has been checked explicitly at NNLO.

The result of the combination of one piece of the SF and the collinear correlator ($\D$) is  free from rapidity divergences and hence can be considered as a valid hadronic quantity. 
For the unpolarized TMDFF in coordinate space we have
\begin{align}
\label{eq:TMDFF}
&\tilde D_{i\to h}(z,\mathbf{L}_\m,\mathbf{l}_{\z_D}) = 
\tilde{\D}_{i\to h}^{(0)}(z,\mathbf{L}_\m,\mathbf{L}_{\sqrt{\d^+\bp^-}})\,
\tilde{S}^\frac{1}{2}({\mathbf L}_\m,{\mathbf L}_{\d^+/\n}) 
\nn\\
&\qquad
=
\tilde{\D}_{i\to h}(z,\mathbf{L}_\m,\pmb\l_{\d^-})\, 
\tilde{S}^{-\frac{1}{2}}({\mathbf L}_\m,{\mathbf L}_{\n\d^-})
\,,
\end{align}
where we have introduced the shorthand notation 
$$\mathbf{l}_{X} \equiv \ln(\m^2/X)
\,,\qquad 
\pmb\l_{\d^-} \equiv \ln(\d^-/\bp^-)\,.$$
In this equation $\tilde{\D}_{i\to h}$ represents the naively calculated collinear matrix element, with no subtraction of the overlapping with the soft region.
If the hard scale in the process, say the mass of the virtual photon in $e^+(p)e^-(\bp)\rightarrow 2\; hadrons$, is given by $Q^2$, then $\z_F$ and $\z_D$ are fractions of $Q^2$, satisfying $\z_F\z_D=Q^4$, where $\z_F=(p^+/\n)^2$ and $\z_D=(\bp^-\n)^2$ (in the following we omit the subscripts $F,\,D$ where unnecessary).
At small values of the impact parameter $\mathbf b$ the renormalized TMDFF can be factorized again in
\begin{align}\label{eq:matching}
\tilde D_{i\to h}(z,{\mathbf L}_\m,\mathbf{l}_{\z}) &=
\int_z^1 \frac{d \tau}{\tau^{3-2\eps}} 
C_{i\to j}\Big(\frac{z}{\tau},{\mathbf L}_\m,\mathbf{l}_{\z}\Big) 
\nn\\
&\times
d_{j\to h}(\tau,\mu)
\,,
\end{align}
where $d_{i\to h}(\xi,\mu)$ is the renormalized integrated FF. 
In Eq.~(\ref{eq:matching}) and in the rest of this letter, the repeating flavor index implies summation. 
The outcome of this work is the calculation at NNLO of the non-singlet and singlet part of quark to quark and quark to anti-quark coefficients, respectively 
$C_{q\to q}(z,{\mathbf L}_\m,\mathbf{l}_{\z})$,   
$C_{q_i\to q_j}(z,{\mathbf L}_\m,\mathbf{l}_{\z})$ and  
$C_{q\to \bar q}(z,{\mathbf L}_\m,\mathbf{l}_{\z})$.

{\bf Regularization.} 
The choice of the infrared (IR) and rapidity regularization scheme is one of the central points for the evaluation of TMDs. 
The regularization should satisfy several important demands, such as: it should respect the exponentiation property of Wilson lines (which is necessary for the existence of the relation in
Eq.~(\ref{eq:splitting})); it should match the singularities of the naively calculated collinear matrix element in the soft limit with the ones of the SF (which is
necessary for a straightforward treatment of the zero-bin subtraction, see footnote \ref{zb_footnote}). 
Additionally, the chosen regularization scheme should be convenient for multi-loop integral computations.

One of the popular choices of regularization is to use tilted Wilson lines, see
e.g. Refs~\cite{Aybat:2011zv,Bacchetta:2013pqa}. 
However, with this regularization the number of loop-integrals and their difficulty is significantly higher than with others. 
The analytical regulator, that was used in NNLO calculation in Refs.~\cite{Gehrmann:2012ze,Gehrmann:2014yya}, is highly efficient for computation and satisfies the necessary requirements, but it is not capable of regularizing rapidity divergences of the SF, which is crucial in the proper definition of an individual TMD.

Here we regularize the rapidity divergences with the $\delta$-regularization, that has been used for the same purpose by many authors, see e.g. Refs.~\cite{Echevarria:2014rua,GarciaEchevarria:2011rb,Cherednikov:2008ua,Vladimirov:2014aja}. 
To regularize the rest of UV and IR divergences we use standard dimensional regularization (DR) with $D=4-2\eps$, while the incoming/outgoing partons are on-shell and massless~\footnote{For renormalization we use $\overline{\text{MS}}$-scheme with the rescaling factor $(4\pi e^{\gamma_E})^\epsilon$.}.

To match the required demands at multi-loop level the $\delta$-regularization is here modified. 
First, in order to supply the non-abelian exponentiation property~\cite{Gatheral:1983cz,Frenkel:1984pz}, and hence the relation in \eq{eq:splitting}, the $\delta$-regulator should be implemented at the operator level, see e.g. the discussion in Ref.~\cite{Vladimirov:2015fea}. 
We thus modify the definition of Wilson lines as~\footnote{We should mention that the presented regulator has some inconveniences typical of such regularizations. 
One of them is the potential violation of gauge invariance. 
However, artificial gauge violating terms can be easily traced and discarded. 
Another inconvenience is that $\delta$ parameter regularizes not only rapidity divergences but also some other soft divergences. 
In general, this is not a problem, since all soft divergences cancel in the final result. Nonetheless, we made a complete analytical calculation, where different soft divergences have different signature, and checked the cancellation individually for every sector. 
The details will be presented in Ref.~\cite{Echevarria:2016scs}.}
\begin{eqnarray}\label{eq:reg}
\tilde W_{\bar n}(0)&=&
P\exp\left[ -i g \int_0^\infty d\sigma A_{-} (\sigma n)\right]
\nn\\
&\rightarrow& 
P\exp\left[ -i g \int_0^\infty d\sigma
A_{-} (\sigma n)e^{-\d^{-}\sigma}\right]
\,,
\nn\\
\tilde S_{\bar n}(0)&=&P\exp\left[ -i g \int_0^\infty d\sigma A_{+} (\sigma \bar n)\right]
\nn\\
&\rightarrow&
P\exp\left[ -i g \int_0^\infty d\sigma
A_{+} (\sigma \bar n)e^{-\d^{+}\sigma}\right]\nn
\,,\\
 S_{n}(0)&=&P\exp\left[ i g \int^0_{-\infty} d\sigma A_{-} (\sigma n)\right]
\nn\\
&\rightarrow&
P\exp\left[ i g \int^0_{-\infty} d\sigma
A_{-} (\sigma  n)e^{+\d^{-}\sigma}\right]
\,,
\end{eqnarray}
where $\delta^\pm \to 0^+$. 
Second, in order to match the IR soft singularities of the naively calculated collinear matrix element and the SF, the $\delta$ in $\D_{i\rightarrow h}$ should be rescaled by $z$, i.e. $\delta\to \delta/z$. 

Such modified regularization is appropriate for being used in multi-loop calculations and for the evaluation of the relevant matrix elements separately.

{\bf Extraction of the matching coefficient.}
In order to extract the matching coefficient at NNLO one needs to evaluate the SF  and the collinear matrix element in Eq.~(\ref{eq:pure_collinear_def}) partonically at NLO and NNLO. 
The obtained functions, together with the renormalization multipliers should be combined into the partonic TMDFF, Eq.~(\ref{eq:TMDFF}). 
The partonic TMDFF then is matched onto the integrated FF in the operator product expansion sense.

At one-loop order the procedure is presented e.g. in Ref.~\cite{Echevarria:2014rua}. 
The complete expression for TMDFF reads
\begin{eqnarray}
\label{eq:TMDFF1L} \tilde D^{[1]}_{q\to q}&=&\tilde \Delta^{[1]}_{q\to q}-\frac{\tilde S^{[1]}}{2}-Z_2^{[1]}+Z_D^{[1]},
\end{eqnarray}
where $Z_2$ is the quark wave-function renormalization constant, and $Z_D$ is the TMDFF operator renormalization constant. 
Note that in this expression, as well as in Eq.~(\ref{CoefFunctStructure}), zero-bin subtractions are explicitely taken into account (the SF is subtracted instead of added). 
Throughout the paper we use superscripts in square brackets to denote the order in the perturbative expansion, e.g. $S=\sum_n a_s^n S^{[n]}$, where $a_s=\frac{g^2}{(4\pi)^2}$ and also the shorthand $\bar z=1-z$.

The rapidity divergences appear in both $\Delta^{[1]}$ and $S^{[1]}$, but cancel in \eq{eq:TMDFF1L}. 
The ultraviolet divergences are renormalized by $Z_2$ and the suitably chosen $Z_D$. 
Therefore $\tilde D^{[1]}_{q\to q}$ is a function of $z$, $\mathbf{L}_\mu$, $\mathbf{l}_{\z}$ and $\epsilon$, which regularizes the IR collinear divergences. 
The collinear divergences are part of the integrated FF, while the matching coefficient $C$ is given by
\begin{align} \tilde C^{[1]}_{i\to j}=\tilde D^{[1]}_{i\to j}-\frac{d^{[1]}_{i\to j}}{z^{2-2\eps}}.
\end{align}
At one-loop order we obtain the well known result~\cite{Collins:2011zzd,Echevarria:2014rua}
\begin{align}
\tilde C^{[1]}_{q\to q}=&
\frac{C_F a_s}{z^2}\Big[
-2\mathbf{L}_{\m/z}{\cal P}_{q\to q}(z)+2 \bar z 
\nn\\ 
& 
+\delta(\bar z)\(-\mathbf{L}_\m^2
+2\mathbf{L}_\m \mathbf{l}_\z
+3\mathbf{L}_\m-\frac{\pi^2}{6}\)\Big]
\,,
\end{align}
and the trivial result $\tilde C^{[1]}_{q\to \bar q }=0$. Here, ${\cal P}_{q\to q}(z)=((1+z^2)/\bar z)_+$ is the quark splitting function.
The plus-distribution is defined as
$
\(f(z)\)_+=f(z)-\delta(\bar z)\int_0^1 dy f(y).
$

At two-loop level the TMDFF is
\begin{eqnarray}
\label{CoefFunctStructure} \tilde D^{[2]}_{i\to j}&=& \tilde \Delta^{[2]}_{i\to j}- \frac{ \tilde S^{[1]} \tilde \Delta^{[1]}_{i\to j} }{2}
-\frac{ \tilde S^{[2]}\d_{ij}}{2} +\frac{3 \tilde S^{[1]}\tilde S^{[1]}}{8}\d_{ij}\nn \\ &&
+\(Z_D^{[1]}-Z_2^{[1]}\)\(\tilde \Delta^{[1]}_{i\to j}-\frac{\tilde S^{[1]}_+\d_{ij} }{2}\)\\
&&\nn+\(Z_D^{[2]}-Z^{[2]}_2-Z_2^{[1]}Z_D^{[1]}+Z_2^{[1]}Z_2^{[1]}\)\d_{ij}.
\end{eqnarray}
The two-loop rapidity divergences appear only in the first line of Eq.~(\ref{CoefFunctStructure}). Notice that in contrast to NLO, where all rapidity divergences
arise with $\delta(\bar z)$ prefactor and cancel trivially between SF and $\Delta$, at NNLO the rapidity divergences arise with an involved
$z$-dependent structure. At two-loop level the rapidity divergences of $\Delta^{[2]}$ and $S^{[2]}$ mix up with UV divergences, and the mixture cancels
in the  combination in Eq.~(\ref{CoefFunctStructure}). In general, Eq.~(\ref{CoefFunctStructure}) possesses a complex system of
cancellations of various divergences~\cite{Echevarria:2016scs}. The realization of all this
cancellation represents an important check of our calculation.

The matching coefficient at two-loop level is given by the combination
\begin{align}\tilde
C^{[2]}_{i\rightarrow f}=\tilde D^{[2]}_{i \rightarrow f}-\tilde C^{[1]}_{i \rightarrow k}\otimes  \frac{d^{[1]}_{k\rightarrow f}}{z^{2-2\epsilon}}-\frac{d^{[2]}_{i \rightarrow f}}{z^{2-2\eps}}
\,,
\end{align}
where the symbol $\otimes$ denotes Mellin convolution in the Bjorken variable $z$, while $k$ is a flavor index.  
Clearly, each addend of this sum is free of rapidity divergences.

{\bf Renormalization group and matching.} 
The renormalization group equations of the TMDFF and the integrated
FF provide also important checks for our calculation. 
We have that
\begin{align}
\label{eq:mevolution}
\m^2\frac{d}{d\m^2}\tilde D_{i\to h}&= 
\frac{1}{2} \g^i_{D}\,\tilde D_{i\to h}
\nn\\
\g_{D}&=
\G^i_{cusp}\mathbf{l}_\z - \g^i_{V}
\,,
\end{align}
where $\G_{cusp}$ is the cusp anomalous dimension, e.g. Ref.~\cite{Echevarria:2012js}.
The result for $\g^q_V$ is extracted from the calculation of the non-singlet part of
the quark form factor~\cite{Moch:1999eb}. 
Then we have
\begin{align}
\label{eq:zevolution}
\zeta\frac{d}{d\zeta}\tilde D_{i\to h}&=
-{\cal D}^i \tilde D_{i\to h}
\,,\quad
2\m^2\frac{d}{d\m^2}{\cal D}^i = \G^i_{cusp}
\,,
\end{align}
which allows the resummation of the rapidity logarithms.
Putting together Eqs.~(\ref{eq:matching}),~(\ref{eq:mevolution}) and~(\ref{eq:zevolution}), one finds
\begin{align}
\zeta\frac{d}{d\zeta}\tilde C_{i\to j}&=-{\cal D}^i\,\tilde  C_{i\to j}\nn\\
\m^2\frac{d}{d\m^2}\tilde C_{i\to j}&=\tilde C_{i\to k}\otimes {\cal K}^i_{k\to j}
\,,
\end{align}
where the convolution is understood in the Bjorken variable $z$ and
\begin{align}
{\cal K}^i_{k\to j}(z)= 
\frac{\delta_{kj}}{2}(\G^i_{cusp}\mathbf{l}_\z
-\g^i_{V})\delta(\bar z)-\frac{{\cal{P}}_{k\to j}(z)}{z^2}.
\end{align}
The function ${\cal{P}}_{i\to j}(z)$ is the DGLAP kernel for the integrated FF at NNLO (see e.g. Refs.~\cite{Moch:2004pa,Vogt:2004mw,Ritzmann:2014mka}).

The evolution equations allow one to write the Wilson coefficient in a more compact form:
\begin{align}\label{eq:C=cal_C}
\tilde C_{i\to j}=
\exp\big[-{\cal D}^i \mathbf{L}_{\sqrt{\zeta}}\big]\, \tilde {\cal C}_{i\to j}
\,.
\end{align}
The most general structure of  $\tilde {\cal C}_{ij}$ of the $n$-th perturbative order is
\begin{align}
\tilde {\cal C}_{ij}^{[n]}&=\sum_{k=0}^{2 n }\tilde {\cal C}_{ij}^{(n;k)}\mathbf{L}_\mu^{k}
\,.
\end{align}
The coefficients $\tilde{\cal C}^{(n;k)}$ are related by the recursive relation
\begin{eqnarray}\label{eq:RGE_recursive}
&&(k+1) \tilde {\mathcal{C}}_{i\to j}^{(n; k+1)} = 
\sum_{r=1}^n\frac{\G_{cusp}^{[r]}}{2}\tilde{\mathcal{C}}_{i\to j}^{(n-r; k-1)}
\\&&\nn
- \frac{\g_{V}^{i[r]}-2(n-r)\beta^{[r]}}{2} \tilde{\mathcal{C}}_{i\to j}^{(n-r; k)}-\tilde{\mathcal{C}}_{i\to k}^{(n-r;k)}\otimes\frac{\mathcal{P}_{k\to j}^{[r]}}{z^2}
\,.
\end{eqnarray}
Thus, given the expressions for the anomalous dimensions one needs only the boundary coefficients $\tilde{ \cal{C}}^{(n,0)}$ in order to reproduce
the complete expression for the matching coefficient. 
In our calculation we evaluate the complete logarithmical structure of the TMDFF and explicitly confirm the relations in Eqs.~(\ref{eq:C=cal_C}) and~(\ref{eq:RGE_recursive}), thus providing a strong check for the whole calculation.

{\bf Results.}
For completeness we present LO and NLO expressions for boundary conditions. They are
\begin{eqnarray}
\tilde{\mathcal{C}}_{q\to q}^{(0;0)}&=&\delta(\bar z),
\\ 
\tilde{\mathcal{C}}_{q\to q}^{(1;0)}&=&\frac{C_F}{ z^2}\Big[\(4p(z) \ln z +2\bar
z\)_+ +
\delta(\bar z)\(6-\frac{3}{2}\pi^2\)\Big]
\nn\,,
\end{eqnarray}
where $p(z)=\frac{1+z^2}{1-z}$. The corresponding quark-antiquark coefficients are zero.

The NNLO coefficient can be decomposed as
\begin{widetext}
\begin{align}
\tilde{\mathcal{C}}_{q\to q}^{(2;0)}(z) &= 
C_F^2 Q_{F}(z) + C_F C_A Q_A(z) + C_F T_R N_f Q_N(z)
\,, \qquad
\tilde{\mathcal{C}}_{q\to \bar q}^{(2;0)}(z) =
C_F\(C_F-\frac{C_A}{2}\) Q_{q\bar q}(z)
\,.
\end{align}
Then the functions $Q_i$ are
\begin{align}
\nn
Q_F(z)&=\frac{1}{z^2}\Bigg\{p(z)\Bigg[40 \Li_3(z)-4\Li_3(\bar z)+4\ln \bar z\Li_2(\bar z)-16\ln z\Li_2(z)
-\frac{40}{3}\ln^3 z+ 18 \ln^2 z\ln\bar z-2\ln^2 \bar z \ln z +\frac{15}{2}\ln^2 z
\\ \nn &
-\(8+\frac{4}{3}\pi^2\) \ln z-40 \zeta_3\Bigg]+\bar z\Bigg[24\Li_2(z)+28 \ln z \ln \bar z+10-\frac{13}{3}\pi^2\Bigg]+\frac{11}{3}(1+z)\ln^3 z
\\&
-\frac{59-9z}{2}\ln ^2 z+2\ln\bar z+(46z-38)\ln z\Bigg\}_++\delta(\bar z)\(-\frac{213}{8}-5\pi^2-12\zeta_3+\frac{397\pi^4}{360}\),
\label{eq:QF}
\end{align}
\begin{align}
\nn
Q_{A}(z)&=
\frac{1}{z^2}\Big\{p(z)\Big[12\Li_3(z)+4\Li_3(\bar z)-4 \ln \(\frac{\bar z}{z^2}\)\Li_2(\bar z)+3\ln^3 z+4\ln \bar z\ln^2 z
\\ &\nn
-\frac{11}{6}\ln^2z+
\frac{10(7-\pi^2)}{3}\ln z+2\zeta_3-\frac{404}{27}\Big]+\bar z\(4\Li_2(\bar z)-\frac{\pi^2}{3}+\frac{44}{3}\)
\\ &
+(8+2z)\ln^2z-2\ln \bar z+\(\frac{116}{3}-\frac{74 z}{3}\)\ln z\Big\}_+
+\delta(\bar z)\(\frac{6353}{81}-\frac{443\pi^2}{36}-\frac{278}{9}\zeta_3+\frac{91\pi^4}{90}\),
\end{align}
\begin{align}
Q_{N}(z)    &=\frac{1}{z^2}\[\(\frac{2}{3}\ln^2 z-\frac{20}{3}\ln z+\frac{112}{27}\)p(z)-\frac{16}{3}\bar z\ln z-\frac{4}{3}\bar z\]_+
+\delta(\bar z)\(-\frac{2717}{162}+\frac{25\pi^2}{9}+\frac{52}{9}\zeta_3\),
\\
\nn
Q_{q\bar q}(z)&=\frac{1}{z^2}\Bigg\{p(-z)\Big[8\Li_3(-z)+16\Li_3(z)-16\Li_3\(\frac{1}{1+z}\)+8\ln z(\Li_2(-z)-\Li_2(z))
\\ &\nn
-6\ln^3 z+\frac{8}{3}\ln^3 (1+z) +12 \ln^2 z\ln(1+z) -\frac{4\pi^2}{3}\ln(1+z)+ 4\zeta_3\Big]
\\ &\nn
-8\bar z \Li_2(\bar z)
+8(1+z)\(\Li_2(-z)+\ln z\ln(1+z)+\frac{\pi^2}{12}\)
\\ &
-8(2+z)\ln^2z+(-38+10 z)\ln z-30\bar z\Bigg\}_+
+\delta(\bar z)\(\frac{187}{4}-6\pi^2-30\zeta_3+\frac{31\pi^4}{45}\).
\end{align}

For the 2-loop singlet part we obtain
\begin{align}
\tilde{\cal C}^{(2;0)}_{q_i\to q_j}&=C_F T_R N_f \left[ \frac{8 \Li_2(\bar z)}{3z^3}\(2(1-z^3)-3z\bar z\)+\frac{22(1+z)\ln^3 z}{3 z^2}+\(\frac{32}{3}-8z^3-11z(1+z)\)\frac{\ln^2 z}{z^3}\right.\nn
\\ &\left.+\frac{4\ln z}{9z^3}(12-174 z-51 z^2-32 z^3)+\frac{2}{27 z^3}(-148-711 z+423 z^2+436 z^3)\right].
\end{align}

\end{widetext}
These expressions represent the main result of this letter.

{\bf Conclusions.}
TMDs are defined as the product in coordinate space of the collinear matrix element and the square root of the soft function.
In this paper we provide the explicit check of this statement for the first time at NNLO, for the quark TMDFF.
The calculation (performed within standard QCD and in Feynman gauge) includes the independent computation of the soft function and the collinear matrix element, and their subsequent recombination into a well-defined TMD.  
We have reformulated the IR and rapidity regularization of Ref.~\cite{Echevarria:2014rua} in order to extend the definition of an individual TMD to multi-loop level. 
We obtain the complete analytical expression for the TMDFF, and comprehensively
investigate the structure of soft/rapidity singularities and their cancellation. 
The cancellation of singularities provides a strong check of the final result. 
As a further check we find a complete agreement between the logarithmical part of the final result and the known predictions of renormalization group. 
The soft factor that has been evaluated in this work, is universal and spin-independent, and thus can be used for the calculation of all TMDs at NNLO.
Finally, the calculation of the TMDFF performed in this work allows us to extract the relevant perturbative matching coefficient at NNLO, necessary to perform the resummation of large logarithms at NNNLL, pushing the phenomenology a step forward.
The applied method can be readily used to obtain other relevant perturbative ingredients.
The detailed report, including the other flavor parton contributions, the explicit expressions, as well as the description of the calculation, will be given in
a separate publication~\cite{Echevarria:2016scs}.

\begin{acknowledgments}
We thank Ahmad Idilbi and Takahiro Ueda for useful discussions. M.G.E. is supported by the ``Stichting voor Fundamenteel Onderzoek der Materie''
(FOM), which is financially supported by the ``Nederlandse Organisatie voor Wetenschappelijk Onderzoek'' (NWO). 
I.S. is supported by the Spanish MECD grant FPA2011-27853-CO2-02 and FPA2014-53375-C2-2-P.
A.V. is supported in part by the European Community-Research Infrastructure Integrating Activity ``Study of Strongly Interacting Matter'' (HadronPhysics3, Grant Agreement No. 283286) and the Swedish Research Council, grants 621-2011-5080 and 621-2013-4287.
\end{acknowledgments}

\newpage

\begin{widetext}
\textbf{Changes in arXiv version 3 compared with version 2}
\\
Equation~\eqref{eq:QF} has been corrected, and it is consistent with Ref.~\cite{Echevarria:2016scs}.

\end{widetext}

\end{document}